\documentclass[pra,showpacs,amsmath,amssymb,preprint,byrevtex]{revtex4} 
     
\usepackage{bm}  
\usepackage{graphicx}
\usepackage{amsmath}  
\newtheorem{Thm}{Theorem}  
\newtheorem{Prop}{Proposition}

\begin{document}

\newcommand{\tr}{\mathop{\mathrm{tr}}\nolimits}
\newcommand{\adj}{\mathop{\mathrm{adj}}}
\renewcommand{\Re}{\mathop{\mathrm{Re}}}
\renewcommand{\Im}{\mathop{\mathrm{Im}}}
\newcommand{\HA}{\mathop{\mathcal{H}}\nolimits}
\newcommand{\D}{\mathop{\mathcal{D}}\nolimits}
\newcommand{\T}{\mathop{\mathcal{T}}\nolimits}
\newcommand{\LA}{\mathop{\mathcal{L}}\nolimits}
\newcommand{\B}{\mathop{\mathcal{B}}\nolimits}
\newcommand{\C}{\mathop{\mathcal{C}}\nolimits}
\newcommand{\SA}{\mathop{\mathcal{S}}\nolimits}
\newcommand{\U}{\mathop{\mathcal{U}}\nolimits}
\newcommand{\V}{\mathop{\mathcal{V}}\nolimits}
\newcommand{\M}{\mathop{\mathcal{M}}\nolimits}
\newcommand{\R}{\mathop{\mathbb{R}}\nolimits}
\newcommand{\CA}{\mathop{\mathbb{C}}\nolimits}
\newcommand{\N}{\mathop{\mathbb{N}}\nolimits}
\newcommand{\I}{\mathop{\mathbb{I}}\nolimits}
\newcommand{\A}{\mathop{\mathcal{A}}\nolimits}

\preprint{}
\title{ The Bloch-vector space for $N$-level systems  \\
--- the spherical-coordinate point of view}
\author{Gen Kimura {\tiny (1,2)}}
\email{gen@hep.phys.waseda.ac.jp}
\author{Andrzej Kossakowski {\tiny (2)}}
\email{kossak@phys.uni.torun.pl}
\affiliation{1 Department of Physics, Waseda University, Tokyo 169--8555, Japan \\ 2 Institute of Physics, Nicolaus Copernicus University, Toru\'n 87--100, Poland}
\begin{abstract}
Bloch-vector spaces for $N$-level systems are investigated from the spherical-coordinate point of view in order to understand their geometrical aspects. We show that the maximum radius in each direction, which is due to the construction of the Bloch-vector space, is determined by the minimum eigenvalue of the corresponding observable (orthogonal generator of $SU(N)$). From this fact, we reveal the dual property of the structure of the Bloch-vector space; if in some direction the space reachs the large sphere (pure state), then in the opposite direction the space can only get to the small sphere, and vice versa. Another application is a parameterization with simple ranges of density operators. We also provide three classes of quantum-state representation based on actual measurements beyond the Bloch vector and discuss their state-spaces. 
\end{abstract}
\maketitle

\section{Introduction}\label{sec:Intro}

One of the most important notions in physics is a state, which includes full information at one moment of the system. In quantum theory, it is thought that the density operator, i.e., a positive operator with a unit trace, gives the representation of a quantum state that includes not only pure states (vector states, or wave functions) but also mixed ones. With a given density operator $\rho$, one will derive any data --- expectation value $\langle A\rangle$ --- of an observable $A = A^\dagger$ through the formula:
\begin{equation}\label{eq:exp}
\langle A \rangle = \tr \rho A. 
\end{equation}
However, since the density operator is not composed of measurable quantities directly available to experimentalists, the question of how to determine the state with experimental data remains as a non-trivial problem. Therefore, it is important to consider another state representation consisting of purely experimental data (actual measurements). 
 
For quantum systems with finite levels, where the dimension of the associated Hilbert space $\HA_N$ is $N < \infty$, the Bloch vector \cite{ref:Bloch,ref:Nielsen,ref:HioeEberly,ref:Lendi,ref:Lendi2,ref:Alicki,ref:Mahler,ref:Jakobczyk,ref:Gen} is one of the candidates that meets the above requirement: Let $\lambda_i$s $ (i=1,\ldots,N^2-1)$ be orthogonal generators of $SU(N)$ (cf. Appendix \ref{sec:GeneratorsOsSUN}) that satisfy 
\begin{equation}\label{eq:generators0} 
\mbox{(i)} \ \lambda^\dagger_i = \lambda_i, \mbox{(ii)} \ \tr \lambda_i = 0, \mbox{(iii)} \ \tr \lambda_i\lambda_j = 2\delta_{ij}.  
\end{equation}
With the identity operator $\I_N$, $\lambda_i$s form an orthogonal basis of the set of all the linear operators with respect to the Hilbert-Schmidt inner product; and hence any density operator $\rho$ can be written in the form:
\begin{eqnarray}\label{eq:Exprho}
\rho &=& \frac{\tr \rho}{N}\I_N + \frac{1}{2}\sum_{i=1}^{N^2-1} (\tr \rho \lambda_i) \lambda_i \nonumber \\
&=& \frac{1}{N}\I_N + \frac{1}{2}\sum_{i=1}^{N^2-1} \langle \lambda_i \rangle \lambda_i, 
\end{eqnarray} 
where use has been made of $\tr \rho = 1$ and $\langle \lambda_i \rangle = (\tr \rho \lambda_i)$ from the formula \eqref{eq:exp}. Physically this means that we only need to know expectation values of $\lambda_i$s to determine the state (density operator). This brings us to another representation of a quantum state by regarding $ \langle \lambda_i \rangle$s themselves as a state; the Bloch vector ${\bm b} \in \R^{N^2-1}$ is defined by a vector in $\R^{N^2-1}$ with components being expectation values of $\lambda_i$s:
\begin{equation}
{\bm b} = (b_1,\ldots,b_{N^2-1}) \equiv (\langle \lambda_1 \rangle,\ldots,\langle \lambda_{N^2-1} \rangle),
\end{equation} 
and from Eq.~\eqref{eq:Exprho} the corresponding density operator $\rho$ is given by the map 
\begin{equation}\label{eq:mapB&Rho}
{\bm b} \to \rho = \frac{1}{N}\I_N + \frac{1}{2}\sum_{i=1}^{N^2-1}b_i \lambda_i.\end{equation} 
 
 While the definition of the Bloch vector is simple as above, it is difficult to know the space (range) of the Bloch vectors; there remains the problem of finding the set of all the Bloch vectors --- the Bloch-vector space $B(\R^{N^2-1})$ --- which is bijectively connected by map \eqref{eq:mapB&Rho} to the set $S(\HA_N)$ of all density operators: 
\begin{equation}
S(\HA_N) \equiv \{ \rho \in \LA(\HA_N) \ : \ (a) \  \rho \ge 0, \  (b)\  \tr \rho = 1\},  
\end{equation}
where $\LA(\HA_N)$ denotes the set of all linear operators on $\HA_N$. So far, there have been many efforts \cite{ref:Mahler,ref:Jakobczyk,ref:Gen,ref:aho,ref:Zyczkowski,ref:Schirmer} to determine $B(\R^{N^2-1})$ for $N$-level systems, and some of its general properties are known: It is a closed convex set in $\R^{N^2-1}$, since $S(\HA_N)$ is a closed convex set and map \eqref{eq:mapB&Rho} is linear homeomorphic. It is a subset of the large ball $D_{r_{l}}(\R^{N^2-1}) \equiv \{{\bm b} \in \R^{N^2-1} : (\sum_{i=1}^{N^2-1} b^2_i)^{\frac{1}{2}} = r_{l} \}$ with radius $r_{l} \equiv \sqrt{\frac{2(N-1)}{N}}$ \cite{ref:Jakobczyk}, which is the minimum ball that includes $B(\R^{N^2-1})$. All the pure states are on the surface of this ball and the states on it are pure \cite{fn:1}; however the points on the surface are not necessarily physical states. On the other hand, $B(\R^{N^2-1})$ includes the small ball $D_{r_{s}}(\R^{N^2-1})$ with radius $r_{s} \equiv \sqrt{\frac{2}{N(N-1)}}$ \cite{ref:Kossakowski}, which is the maximum ball included in $B(\R^{N^2-1})$ \cite{fn:zyc}:
\begin{equation}\label{eq:includion}
D_{r_s}(\R^{N^2-1}) \subseteq B(\R^{N^2-1}) \subseteq D_{r_l}(\R^{N^2-1}).
\end{equation} 
Only in $2$-level systems ($N=2$) does it follow that $r_s = r_l \ ( = 1)$ and the Bloch-vector space comes to be a ball, which is well known as the Bloch ball \cite{ref:Bloch,ref:Nielsen,ref:Mahler}. However, the inclusive relations \eqref{eq:includion} do not coincide when $N \ge 3$, and $B(\R^{N^2-1})$ has not been determined until quite recently: In Ref.~\cite{ref:Gen}, $B(\R^{N^2-1})$ was analytically determined for arbitrary $N$-level systems, clarifying the origin of the different structures between $2$-level and $N$-level systems ($N\ge 3$). From the physical point of view, the determination gives a theoretical prediction of the range of data that will be measured in experiment, and the analytic forms can be used for such a purpose. However, unlike the $2$-level case, the forms in higher-level systems are quite complex, and it is still hard to understand the geometrical character of the space, although the geometrical knowledge of state space is quite useful for comprehending global aspects, in connection with experiments, of not only the state but also the dynamics. Therefore, it is still worth trying to understand the Bloch-vector space, especially its geometrical character, which gives us an overall picture of the space. 

In this Letter, we will investigate the structure of the Bloch-vector space with a different strategy from that of Ref.~\cite{ref:Gen}, namely from the spherical-coordinate point of view. We reveal that not a maximum but a minimum spectrum of orthogonal generators of $SU(N)$ of each direction determines the space (Theorem \ref{Thm:Main} in Sec.~\ref{sec:RadiusOfBS}). This provides clear perception about a geometry of the space and, as one of the applications, we will show a dual property of the Bloch-vector space (Theorem \ref{eq:Dual} in Sec.~\ref{sec:dual}): if in some direction the space reaches the surface of large ball $D_{r_{l}}(\R^{N^2-1})$ (pure state), then in the opposite direction it can only reach the surface of small ball $D_{r_{s}}(\R^{N^2-1})$, and vice versa. As an another application, we show that spherical coordinate of the Bloch vector could be a useful parameterization of density operators (Sec.~\ref{sec:para}) for the purpose of numerical experiments. In the final section (Sec.~\ref{sec:dis}), we discuss state-representations based on actual measurements beyond the Bloch-vector representation. We provide three classes (C1), (C2), and (C3), where (C3) is that of the Bloch vector, clarifying its physical and mathematical reasons in choosing special observables to represent quantum state. We also discuss their state spaces and show Theorem \ref{Thm:Main} still holds for class (C2). This fact will be helpful in the search for more useful state-representation than that of the Bloch vector.

\section{Toward the spherical-coordinate point of view}

We begin by recalling a basic property in probability theory that an expectation value takes its value between minimum and maximum values of the random variable. This applies also in quantum theory where spectra of an observable determine its range. In the meantime, since components $b_i\ (i=1,\ldots,N^2-1)$ of the Bloch vector are expectation values of orthogonal generators $\lambda_i\ (i=1,\ldots,N^2-1)$, their ranges are restricted between minimum eigenvalue $m(\lambda_i)$ and maximum eigenvalue $M(\lambda_i)$ of $\lambda_i$:
\begin{equation}\label{eq:RangeByNorm}
m(\lambda_i) \le b_i \le M(\lambda_i). 
\end{equation}
This simple fact might give information on the Bloch-vector space, and indeed the space should be restricted as 
\begin{equation}\label{eq:LooseRes}
B(\R^{N^2-1}) \subseteq \{{\bm b} \in \R^{N^2-1} : m(\lambda_i) \le b_i \le M(\lambda_i) \}. 
\end{equation}
However, this restriction never determines the space exactly --- the inclusive relation is always proper. Let us explain this situation in a $2$-level system, where the Bloch-vector space $B(\R^{3})$ corresponds to the Bloch ball with radius $1$: The spectrum of the generators $\{\lambda_i\}_{i=1}^3$ (Pauli's spin operators) are $\pm 1$, i.e., $m(\lambda_i) = -1,\ M(\lambda_i) = 1$, and Eq.~\eqref{eq:LooseRes} comes to be
\begin{equation}\
B(\R^{3}) = D_1(\R^{3}) \subseteq \{{\bm b} \in \R^{3} : -1 \le b_i \le 1 \}, 
\end{equation}
where the right-hand side gives not the ball but a cube. One should consider why the above consideration was not enough to determine the Bloch-vector space. Generally speaking, when an expectation value achieves the maximum (or minimum) eigenvalue, the state should be the eigenstate of the corresponding observable. However, since not all the generators $\lambda_i$ are commutative with each other, the cases where the equalities in Eq.~\eqref{eq:RangeByNorm} of non-commutative $\lambda_i$'s hold are prohibited by the principle of quantum mechanics --- non-commutative observables do not generally have simultaneous eigenstates. 

However, we can further proceed to investigate the space using the same philosophy. Although in the above discussion, the expectation values are only considered for each $\lambda_i \ (i=1.\ldots,N^2-1)$, one can consider the same restriction \eqref{eq:RangeByNorm} for $\lambda_{\bm n} \equiv \sum_{i=1}^{N^2-1} n_i\lambda_i$ in all directions ${\bm n} \in \R^{N^2-1} \ (\sum_{i=1}^{N^2-1} n^2_i = 1)$. This can be done by considering the spherical-coordinate of the Bloch vector; ${\bm b} = r {\bm n} \ (r \ge 0)$ in each direction ${\bm n} \in \R^{N^2-1} \ (\sum_{i=1}^{N^2-1} n^2_i = 1)$, where the corresponding density operator is 
\begin{equation}\label{eq:DOinSC}
\rho = \frac{1}{N}\I_N + \frac{1}{2}\sum_{i=1}^N (r n_i) \lambda_i = \frac{1}{N}\I_N + \frac{1}{2} r \lambda_{\bm n}.
\end{equation}
Note that the radius $r$ of the Bloch vector is the expectation value of $\lambda_{\bm n}$, i.e., $r = \tr \rho \lambda_{\bm n}$, since $\tr \lambda_{\bm n} = 0$ and $\tr \lambda_{\bm n}^2 = 2$. By applying the basic property of the expectation value, $r$ is bounded above by the maximum eigenvalue $M(\lambda_{\bm n})$ of $\lambda_{\bm n}$ \cite{fn:2}:  
\begin{equation}
r \le M(\lambda_{\bm n}).  
\end{equation}
This provides us stronger restriction of the Bloch-vector space than that achieved by Eq.~\eqref{eq:LooseRes}: 
\begin{equation}\label{eq:BisSubsetByM}
B(\R^{N^2-1}) \subseteq \{{\bm b} = r{\bm n} \in \R^{N^2-1} : r \le M(\lambda_{\bm n}) \}. 
\end{equation}
This is of importance especially in a $2$-level system; since every $\lambda_{\bm n}$ has eigenvalues $\pm 1$ independent of the direction ${\bm n}$ (see Sec.~\ref{sec:EVofOG} for details) and hence $M(\lambda_{\bm n}) = 1 $, Eq.~\eqref{eq:BisSubsetByM}, perhaps surprisingly, determines the Bloch-vector space (Bloch ball) exactly:
\begin{equation}
B(\R^{3}) = D_1(\R^{3}) = \{{\bm b} = r{\bm n} \in \R^{3}  : r \le M(\lambda_{\bm n}) = 1\}. 
\end{equation}

Unfortunately, this simple discussion still does not determine the Bloch-vector space for $N$-level systems ($N \ge 3$), as is shown in Sec.~\ref{sec:dual}. This fact, on first sight, tells us that the information of the spectra of generators $SU(N)$ is generally not enough to determine the Bloch-vector space. However, we will show (in Theorem \ref{Thm:Main} in Sec.~\ref{sec:RadiusOfBS}) that the Bloch-vector space is completely determined by the spectra --- by not the maximum but the minimum eigenvalue $ m(\lambda_{\bm n})$ of generator $\lambda_{\bm n}$ --- in the form  
\begin{equation}
B(\R^{N^2-1}) = \{{\bm b} = r{\bm n} \in \R^{N^2-1} : r \le \frac{2}{N | m(\lambda_{\bm n})|} \}, 
\end{equation}
instead of Eq.~\eqref{eq:BisSubsetByM}. This is nothing but a construction of the Bloch-vector space from the spherical-coordinate point of view.

\section{The radius of the Bloch-vector space}\label{sec:RadiusOfBS}

Let us again consider the spherical coordinate of the Bloch vector; ${\bm b} = r {\bm n} \ (r \ge 0)$ in each direction ${\bm n} \in \R^{N^2-1} \ (\sum_{i=1}^{N^2-1} n^2_i = 1)$, where the corresponding density operator is given by Eq.~\eqref{eq:DOinSC}. Since $B(\R^{N^2-1})$ is a closed bounded convex set that includes origin ${\bm b} = {\bm 0}$ \cite{fn:3}, there exists the maximum radius $||{\bm b}_{\bm n}||_{\max}$ in each direction ${\bm n} \in \R^{N^2-1}$: 
\begin{equation}\label{def:bmax}
||{\bm b}_{\bm n}||_{\max} \equiv \max_{ r{\bm n} \in B(\R^{N^2-1}) } \{r \}, 
\end{equation}
with which the Bloch-vector space is given by 
\begin{equation}
B(\R^{N^2-1}) = \{{\bm b} = r{\bm n} \in \R^{N^2-1} : r \le ||{\bm b}_{\bm n}||_{\max} \}.  
\end{equation}
From relations \eqref{eq:includion}, we know the restrictions:  
\begin{equation}\label{eq:restrictionofbmax}
r_s = \sqrt{\frac{2}{N(N-1)}} \le ||{\bm b}_{\bm n}||_{\max} \le \sqrt{\frac{2(N-1)}{N}} = r_l. 
\end{equation}
On the other hand, from Eq.~\eqref{eq:BisSubsetByM}, $||{\bm b}_{\bm n}||_{\max}$ is also bounded above by
\begin{equation}\label{eq:BAbyMax}
 ||{\bm b}_{\bm n}||_{\max} \le M(\lambda_{\bm n}).
\end{equation}
However, we reveal that it is not maximum eigenvalue but minimum eigenvalue that determines $||{\bm b}_{\bm n}||_{\max}$, hence also the Bloch-vector space:
\begin{Thm}\label{Thm:Main}
Let $\lambda_{\bm n} \equiv \sum_{i=1}^{N^2-1} n_i \lambda_i$ with direction vector ${\bm n} \in \R^{N^2-1} \ (\sum_{i=1}^{N^2-1} n^2_i =1 )$ and $m(\lambda_{\bm n})$ be its minimum eigenvalue. Then, 
\begin{equation}\label{eq:bmax=||}
||{\bm b}_{\bm n}||_{\max} = \frac{2}{N|m(\lambda_{\bm n})|}. 
\end{equation}
Namely, the Bloch-vector space in the spherical coordinate is given by
\begin{equation}\label{eq:TheBSinSC}
B(\R^{N^2-1}) = \{{\bm b} = r {\bm n} \in \R^{N^2-1} : r \le  \frac{2}{N|m(\lambda_{\bm n})|} \}. 
\end{equation}
\end{Thm}

{\bf Proof}

Before proving this, let us note that operator $\lambda_{\bm n} \equiv {\bm n}\cdot {\bm \lambda}$ for any direction ${\bm n}$ satisfies 
\begin{equation}\label{eqs:lambdan}
\mbox{(i)}\ \lambda_{\bm n}^\dagger = \lambda_{\bm n}, \ \mbox{(ii)} \ \tr \lambda_{\bm n} = 0,\ \mbox{(iii)} \ \tr \lambda^2_{\bm n} = 2, 
\end{equation}
as are easily proved from Eqs.~\eqref{eq:generators0}. From these properties, one obtains 
\begin{equation}\label{eq:inf}
\inf_{| \psi \rangle \in \HA_N \ (||\psi||=1)} \langle \psi|\lambda_{\bm n}|\psi\rangle = m(\lambda_{\bm n}) < 0. 
\end{equation}
(From (i) in Eqs.~\eqref{eqs:lambdan}, all the eigenvalues $a_i \ (i=1,\ldots,N)$ of $\lambda_{\bm n}$ are real, while the corresponding eigenvectors form a complete orthonormal system; hence, the first equality holds \cite{ref:Reed&Simon}. Suppose that $a_N \equiv m({\lambda_{\bm n}}) \ge 0$, then $\tr \lambda_{\bm n} = \sum_{i=1}^N a_i \ge 0$, which contradicts (ii) in Eqs.~\eqref{eqs:lambdan} instead of the case in which all the eigenvalues are 0. However, the latter case implies $\lambda_{\bm n} = 0$, and this contradicts independency among $\lambda_i$s. Hence, we obtain $m(\lambda_{\bm n}) < 0$. Note also that $M(\lambda_{\bm n}) > 0$, which is shown in the same manner). 

In order to prove \eqref{eq:bmax=||}, it is enough to check the range of $r$ where the positivity of $\rho = \frac{1}{N}\I_N + \frac{1}{2}r \lambda_{\bm n} $ is satisfied: 
\begin{equation}
\inf_{|\psi\rangle \in \HA_N \ (||\psi||=1)}\langle\psi | \frac{1}{N}\I_N + \frac{r}{2}  \lambda_{\bm n}|\psi \rangle \ge 0 \Leftrightarrow  \frac{1}{N} + \frac{r}{2} \inf \langle \psi|\lambda_{\bm n}|\psi\rangle \ge 0. 
\end{equation}
From Eq.~\eqref{eq:inf}, this is equivalent to
$$
r \le \frac{2}{N |m_{\bm n}|}. 
$$
Since $||{\bm b}_{\bm n}||$ is the maximum $r$ of this range, Eq.~\eqref{eq:bmax=||} holds.  
\hfill{QED}

 
We expect that Theorem \ref{Thm:Main} gives another comprehensive perspective of the Bloch-vector space; namely, it enables us to capture the Bloch-vector space in the sense that all the information of the space is accumulated in that of the minimum eigenvalues of generators. Although this characterization of the Bloch-vector space has yet to give an explicit determination of the space as achieved in Ref.~\cite{ref:Gen}, there are no difficulties in determining the space in numerical computation (see Appendix \ref{2dimSec}). In the next section, we also provide some useful techniques for obtaining eigenvalues of orthogonal generators of $SU(N)$ and illustrate the result of Theorem \ref{Thm:Main} in $2$-level and $3$-level systems.

\section{Eigenvalues of orthogonal generators of $SU(N)$}\label{sec:EVofOG}
In this section, we provide one of the techniques for solving an eigenvalue-problem of orthogonal generators of $SU(N)$. The equation of eigenvalues for $\lambda_i$ can be written as  
$$
\det (x\I_N - \lambda_i) = \sum_{j=0}^N c_j x^{N-j} = 0, (c_0 = 1), 
$$ 
where coefficients $c_i$s can be derived through the recurrence formula: 
\begin{equation}\label{eq:ExplicitNewtonFormula2}
kc_k 	= \sum_{q=1}^{k}(-1)^{q-1} \tr(\lambda^q_i) c_{k-q} \quad (1\le k \le N),
\end{equation} 
which is called Newton's formula \cite{ref:DescartesAndNewton}. Furthermore, from Eqs.~\eqref{eq:generators} in Appendix \ref{sec:GeneratorsOsSUN}, traces of any numbers of multiplications of generators can be derived by multiplying Eq.~\eqref{eq:multilambda} by generators in sequence and taking a trace of them: 
\begin{equation}\label{eq:lambdalambdalambda}
\tr\lambda_i = 0,\ \tr\lambda_i\lambda_j = 2\delta_{ij}, \ \tr \lambda_i\lambda_j\lambda_k = 2 z_{ijk},\  \tr \lambda_i\lambda_j\lambda_k\lambda_l = \frac{4}{N}\delta_{ij}\delta_{kl} + 2 \sum_{m=1}^{N^2-1}z_{ijm}z_{mkl}, \cdots.
\end{equation}
In particular, it holds that
\begin{equation}\label{eq:lambda^n}
\tr\lambda_i = 0,\ \tr\lambda^2_i= 2, \ \tr \lambda^3_i= 2g_{iii},\  \tr \lambda^4_i = \frac{4}{N} + 2\sum_{m=1}^{N^2-1}g_{iim}g_{mii}, \cdots,
\end{equation}
where use has been made of antisymmetric and symmetric properties of $f_{ijk}$ and $g_{ijk}$. Hence, from the formula \eqref{eq:ExplicitNewtonFormula2}, the coefficients can be obtained as  
\begin{equation}\label{eq:coef}
 c_1 = 0, \ c_2 = -1,\  c_3 = \frac{2}{3} g_{iii}, \ c_4 =  \frac{N-2}{2N} - \frac{1}{2}\sum_{m=1}^{N^2-1}g_{iim}g_{mii}\cdots.
\end{equation}
In the following, we illustrate how one can solve eigenvalue-problems in $2$-level and $3$-level systems as examples: 

[N=2] From Eqs.~\eqref{eq:coef}, an equation of eigenvalues is 
\begin{equation}
\det (x\I_2 - \lambda_i) =  x^2 -1 = (x-1)(x+1).
\end{equation} 
It follows that eigenvalues of $\lambda_i$ are $\pm 1$ independent of the choice of generators; in particular the same holds for $\lambda_{\bm n}$ for any direction. Consequently, from a result of Theorem \ref{Thm:Main} with $N=2$ and $m(\lambda_{\bm n}) = -1$, it is reproduced that the Bloch-vector space is merely a ball (the Bloch ball).  
\begin{figure}
\includegraphics[height=0.3\textwidth]{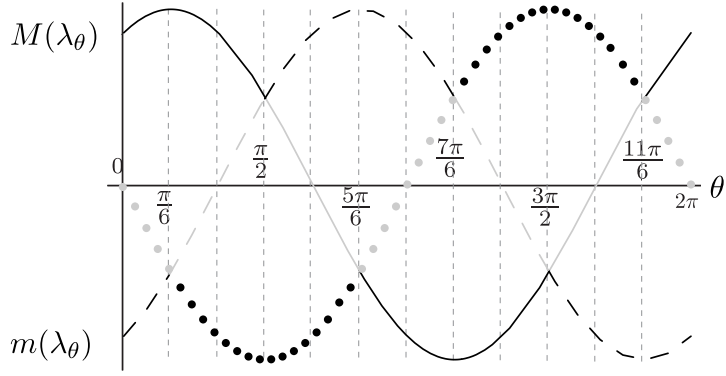}
\caption{The graph of maximum eigenvalue $M(\theta)$ and minimum eigenvalue $m(\theta)$ of $\lambda_\theta$. The solid line, dashed line, and dotted line are for eigenvalues of $\frac{2\sqrt{3}}{3}\sin(\theta + \frac{\pi}{3})$, $-\frac{2\sqrt{3}}{3}\sin\theta$, and $\frac{2\sqrt{3}}{3}\sin(\theta - \frac{\pi}{3})$, respectively.}\label{fig:maxmin}
\end{figure}
 
[N=3] From Eqs.~\eqref{eq:coef}, an equation of eigenvalues is
\begin{equation}
\det (x\I_3 - \lambda_i) =  x^3  - x  -\frac{2}{3}g_{iii}. 
\end{equation}
As an example, consider the Gell-Mann operators \cite{ref:Mahler}, with non-vanishing structure constants $g_{ijk}$: $g_{118} = g_{228} = g_{338} = -g_{888} = \sqrt{3}/3, \ g_{448} = g_{558} = g_{668} = g_{778} = -\sqrt{3}/6, \ g_{146} = g_{157} = g_{256} = g_{344} = g_{355} = -g_{247} =-g_{366}= -g_{377} =  1/2$. Since, $g_{iii} = 0 \ (i=1,\ldots,7)$ and $g_{888} = -\sqrt{3}/3$, the eigenvalues of them are $\{\pm 1, 0\}$ for $i=1,\ldots,7$ and $\{\sqrt{3}/3, -2\sqrt{3}/3\}$ for $i=8$. 


To illustrate Theorem \ref{Thm:Main} in a $3$-level system, we shall obtain one of the $2$-dimensional sections \cite{ref:Mahler,ref:Jakobczyk,ref:Gen} of the Bloch-vector space. For this purpose, let us consider $\lambda_{\theta} \equiv \cos \theta \lambda_i + \sin \theta \lambda_8 \ ( i = 1,2,3) $, where $0 \le \theta < 2 \pi$. Since $g_{iii} = 0, \ g_{ii8} = \sqrt{3}/3, \ g_{i88} =0, \ g_{888} = -\sqrt{3}/3$ ($i=1,2,3$), eigenvalue equation for $\lambda_{\theta}$ is 
\begin{eqnarray}
0= \det (x\I_N - \lambda_\theta) &=& x^3  - x  -\frac{2\sqrt{3}}{9}\sin\theta(3\cos^2\theta - \sin \theta) \nonumber \\
 &=& (x + \frac{2\sqrt{3}}{3}\sin\theta)(x - \frac{2\sqrt{3}}{3}\sin(\theta + \frac{\pi}{3}))(x - \frac{2\sqrt{3}}{3}\sin(\theta - \frac{\pi}{3})).
\end{eqnarray}
Hence the eigenvalues of $\lambda_\theta$ are $\{-\frac{2\sqrt{3}}{3}\sin\theta, \frac{2\sqrt{3}}{3}\sin(\theta \pm \frac{\pi}{3}) \}$; maximum and minimum eigenvalues for each $\theta$ (see Fig.~\ref{fig:maxmin}) are 
\begin{subequations}
\begin{eqnarray}
M(\lambda_\theta) &=& 
\left\{
\begin{array}{ccc}
 \frac{2\sqrt{3}}{3}\sin(\theta + \frac{\pi}{3}) & : & (0\le \theta < \frac{\pi}{2},  \frac{11\pi}{6} \le \theta < 2\pi),\\
 \frac{2\sqrt{3}}{3}\sin(\theta - \frac{\pi}{3}) & : & (\frac{\pi}{2} \le \theta < \frac{7\pi}{6}),\\
 -\frac{2\sqrt{3}}{3}\sin \theta & : & (\frac{7\pi}{6}\le \theta < \frac{11\pi}{6}),\\
\end{array}
\right. \label{eq:Max} \\
m(\lambda_\theta) &=& 
\left\{
\begin{array}{ccc}
 \frac{2\sqrt{3}}{3}\sin(\theta - \frac{\pi}{3}) & : & (0\le \theta < \frac{\pi}{6},  \frac{3\pi}{2} \le \theta < 2\pi),\\
 -\frac{2\sqrt{3}}{3}\sin\theta & : & (\frac{\pi}{6} \le \theta < \frac{5\pi}{6}),\\
 \frac{2\sqrt{3}}{3}\sin (\theta+ \frac{\pi}{3}) & : & (\frac{5\pi}{6}\le \theta < \frac{3\pi}{2}). \\
\end{array}
\right. \label{eq:min}
\end{eqnarray}
\end{subequations}
\begin{figure}
\includegraphics[height=0.4\textwidth]{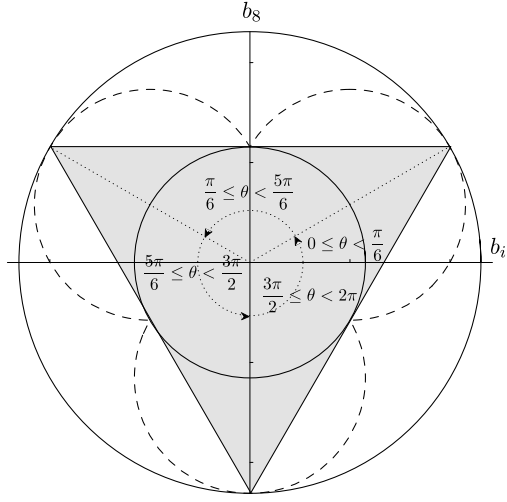}
\caption{The gray region (regular triangle) is a $2$-dimensional section of $b_i \ (i=1,2,3)$ and $b_8$ of the Bloch-vector space $B(\R^8)$ for $3$-level systems, which was drawn using Theorem \eqref{Thm:Main} and Eq.~\eqref{eq:min}; the large and small disks are sections of $ D_{r_l}(\R^{8}) $ and $D_{r_s}(\R^{8})$, respectively (see inclusion relations \eqref{eq:includion}). The broken line, which does not construct the Bloch-vector space, is the maximum eigenvalue in each direction drawn by Eq.~\eqref{eq:Max}. }\label{fig:i8}
\end{figure}
From Eq.~\eqref{eq:bmax=||} and Eq.~\eqref{eq:min}, the maximum radius $||{\bm b}_\theta||_{\max}$ is given by 
\begin{equation}\label{eq:N=3ex}
||{\bm b}_\theta||_{\max} = \frac{-1}{\sqrt{3}\sin(\theta - \frac{\pi}{3})  }, 
\end{equation}
in the range of $0\le \theta < \frac{\pi}{6}$ and $ \frac{3\pi}{2} \le \theta < 2\pi$. Let $b_i \equiv ||{\bm b}_\theta||_{\max} \cos \theta, b_8 \equiv ||{\bm b}_\theta||_{max} \sin \theta$; then Eq.~\eqref{eq:N=3ex} is equivalent to 
\begin{subequations}\label{eq:triangle}
\begin{equation}
b_8 = \sqrt{3}b_i - \frac{2}{\sqrt{3}} \ (0\le \theta < \frac{\pi}{6},\ \frac{3\pi}{2} \le \theta < 2\pi). 
\end{equation}
In the same manner, one obtains 
\begin{equation}
b_8 = \frac{1}{\sqrt{3}} \ (\frac{\pi}{6} \le \theta < \frac{5\pi}{6}), 
\end{equation}
and 
\begin{equation}
b_8 = -\sqrt{3}b_i - \frac{2}{\sqrt{3}} \ (\frac{5\pi}{6}\le \theta < \frac{3\pi}{2}),
\end{equation}
\end{subequations}
which are also obtained in Ref.~\cite{ref:Gen} as an example (See Eq.~(32) in the reference).  

Figure \ref{fig:i8} is a $2$-dimensional section of the Bloch-vector space for $b_i\ (i=1,2,3)$ and $b_8$, which was drawn using Theorem \ref{Thm:Main}, Eq.~\eqref{eq:min}, and Eqs.~\eqref{eq:triangle}. (See also Fig.~\ref{fig:34level} [I] in Appendix \ref{2dimSec}). The grey region (regular triangle) is a $2$-dimensional section of $b_i \ (i=1,2,3)$ and $b_8$ of the Bloch-vector space. The sections of large and small balls $ D_{r_{l}}(\R^{8}) $ and $D_{r_{s}}(\R^{8})$ are also drawn here to illustrate the inclusion relations \eqref{eq:includion}. The sharp bending structure appears due to the changes of minimum eigenvalues in Eq.~\eqref{eq:min} at the angle $\theta = \frac{\pi}{6}, \frac{5\pi}{6}, \frac{3\pi}{2}$ (see Fig.~\ref{fig:maxmin}). The maximum eigenvalues of $\lambda_\theta$ in each direction are also plotted in a broken line using Eq.~\eqref{eq:Max}, which shows that they do not determine the Bloch-vector space. (See inclusion relation \eqref{eq:BisSubsetByM}).  

In Fig.~\ref{fig:i8}, one might notice the dual properties. When in some direction ($\theta = \pi/6, \ 5\pi/6, \ 3\pi/2$ in the figure) the space reachs the surface of the large ball (pure state), then in the opposite direction ($\theta = 7\pi/6, 11\pi/6, \pi/2 $, respectively, in the figure ) the space can get to only the surface of the small ball, and vice versa. In the next section, we show that this property universally holds for any $N$-level system.  

\section{Dual properties of the Bloch-vector space}\label{sec:dual}

In this section, we show the existence of a dual property of the Bloch-vector space. We shall start with some properties of the eigenvalues of the orthogonal generators of $SU(N)$: 
\begin{Prop}\label{Prop:EiSUN}
 Let $a_j$s $ (j=1,\ldots,N)$ be eigenvalues of $\lambda_{\bm n}$ in descending order: $M(\lambda_{\bm n}) \equiv a_1 \ge a_2 \ge \cdots \ge a_{N-1} \ge a_N \equiv m(\lambda_{\bm n})$. Then, the following properties [I], [II], and [III] hold: 

[I] The minimum eigenvalues are restricted by
\begin{equation}\label{ineq:rangeof|m|}
\sqrt{\frac{2}{N(N-1)}} \le |m(\lambda_{\bm n})| \le \sqrt{\frac{2(N-1)}{N}}, \ \end{equation}

[II] If $|m(\lambda_{\bm n})|$ achieves maximum in Eq.~\eqref{ineq:rangeof|m|}, i.e., $|m(\lambda_{\bm n})|= \sqrt{\frac{2(N-1)}{N}}$, then all other eigenvalues are the same: 
\begin{equation}
a_j = \sqrt{\frac{2}{N(N-1)}} \ (j=1,\ldots,N-1). 
\end{equation}

[III] If $|m(\lambda_{\bm n})|$ takes minimum in Eq.~\eqref{ineq:rangeof|m|}, i.e., $|m(\lambda_{\bm n})|= \sqrt{\frac{2}{N(N-1)}}$, then 
\begin{equation}
a_j = - \sqrt{\frac{2}{N(N-1)}} \ (j=2,\ldots,N-1), \ M(\lambda_{\bm n}) = a_1 =  \sqrt{\frac{2(N-1)}{N}}. 
\end{equation}
\end{Prop}
Notice that the same holds for $M(\lambda_{\bm n})$ when one substitutes the maximal values of $M(\lambda_{\bm n})$ instead of $|m(\lambda_{\bm n})|$, since $M(\lambda_{\bm n}) = - m(\lambda_{-{\bm n}})$. 

{\bf Proof of Proposition \ref{Prop:EiSUN}}

\noindent From Eqs.~\eqref{eqs:lambdan}, it follows that 
\begin{equation}\label{eqs:lambdan2}
\mbox{(i)}^\prime \ a_j \in \R, \  \mbox{(ii)}^\prime \  \sum_{j=1}^N a_j = 0, \ \mbox{(iii)}^\prime \ \sum_{j=1}^{N} a^2_j = 2. 
\end{equation}
Note that in a $2$-level system, the only solutions are $a_1=M(\lambda_{\bm n})=1,a_2= m(\lambda_{\bm n})=-1$, which compose the Bloch ball.  In order to find extremal values of $a_i$ satisfying Eqs.~\eqref{eqs:lambdan2}, we define $f_i(a_1,\ldots,a_N;\alpha_i,\beta_i) \equiv a_i + \alpha_i (\sum_{j=1}^N a_j) + \beta_i(\sum_{j=1}^{N} a^2_j - 2)$ and solve equations:
\begin{eqnarray}
\frac{\partial}{\partial a_j} f_i(a_1,\ldots,a_N;\alpha_i,\beta_i) = 0 \ (j=1,\ldots,N) &\Leftrightarrow& \delta_{ij} + \alpha_i + 2\beta_i a_j = 0, \label{eqs:1}\\
\frac{\partial}{\partial \alpha_i} f_i(a_1,\ldots,a_N;\alpha_i,\beta_i) = 0 &\Leftrightarrow& \sum_{j=1}^N a_j = 0, \label{eq:2}\\
\frac{\partial}{\partial \beta_i} f_i(a_1,\ldots,a_N;\alpha_i,\beta_i) = 0 &\Leftrightarrow& \sum_{j=1}^N a^2_j = 2.\label{eq:3} 
\end{eqnarray} 
First, summing Eqs.~\eqref{eqs:1} over $j = 1,\ldots,N$ and considering Eq.~\eqref{eq:2}, one obtains 
\begin{equation}
1 + N \alpha_i = 0 \Leftrightarrow \alpha_i = - \frac{1}{N}. 
\end{equation}
Next, multiplying Eqs.~\eqref{eqs:1} by $a_j$, summing them over $j = 1,\ldots,N$ and considering Eq.~\eqref{eq:3}, one obtains 
\begin{equation}
a_i + 4 \beta_i = 0 \Leftrightarrow \beta_i = - \frac{a_i}{4}. 
\end{equation}
From these, Eqs.~\eqref{eqs:1} become 
\begin{equation}\label{eqs:1new}
 \delta_{ij} -\frac{1}{N}  - \frac{1}{2} a_i a_j = 0\ (j=1,\ldots,N). 
\end{equation}
When $j=i$, we obtain the solutions 
$$
a_i = \pm \sqrt{\frac{2(N-1)}{N}}. 
$$
This means that maximum and minimum values of $a_i$ are $\pm \sqrt{\frac{2(N-1)}{N}}$, since $|a_j|$ is bounded above by $\sqrt{2}$ from $\mbox{(iii)}^\prime$, which assures that $a_j$ has minimum and maximum values; namely it follows 
$$
|m(\lambda_{\bm n})|, \ M(\lambda_{\bm n}) \le \sqrt{\frac{2(N-1)}{N}}. 
$$
While $a_i$ achieves the values, the other elements $a_j \ (j\neq i)$ take values of $$
a_j = \mp \sqrt{\frac{2}{N(N-1)}}\ (j \neq i),
$$ 
which are obtained by taking $j\neq i$ in Eqs.~\eqref{eqs:1new}. This completes the proof of [II]. 

Next, assume that $|m(\lambda_{\bm n})| < \sqrt{\frac{2}{N(N-1)}}$, and we will show the contradiction. Let $k$-numbers ($N-1 \ge  k \ge 1$) of $a_i$s be positive semi-definite, i.e., $a_1 \ge \cdots \ge a_k \ge 0 > a_{k+1} \ge \cdots \ge a_N = m(\lambda_{\bm n}) $. Then, it holds that 
\begin{equation}\label{eq:assum}
|a_j | < \sqrt{\frac{2}{N(N-1)}} \ (j = k+1, \ldots, N). 
\end{equation}
Hence we obtain 
\begin{equation}\label{eq:ineqcontra}
\sum_{j=k+1}^N a^2_j  < \frac{2(N-k)}{N(N-1)} \Leftrightarrow \sum_{j=1}^k a^2_j > 2 - \frac{2(N-k)}{N(N-1)}. 
\end{equation}
Fixing these values of $a_j$s $(j=k+1,\ldots,N)$ temporarily, we define $g(b_1,\ldots,b_k;\gamma) \equiv \sum_{j=1}^k b^4_j + \gamma( \sum_{j=1}^k b^2_j - \sum_{j=k+1}^N |a_j| ) $, where $b_i^2 \equiv a_i \ge 0 \ (i =1,\ldots,k) $ to find the maximum value of $\sum_{j=1}^k a^2_j = \sum_{j=1}^k b^4_j$ and solve equations:
\begin{eqnarray}
\frac{\partial}{\partial b_j} g(b_1,\ldots,b_k;\gamma) = 0 \ (j=1,\ldots,k) &\Leftrightarrow& 4 b^3_j + 2 \gamma b_j = 0, \label{eqs:4}\\
\frac{\partial}{\partial \gamma} g(b_1,\ldots,b_k;\gamma) = 0 &\Leftrightarrow& \sum_{j=1}^k b_j^2 = \sum_{j=k+1}^N |a_j|. \label{eq:5} 
\end{eqnarray}
Notice that from Eq.~\eqref{eq:5} all $|b_i|$s are bounded from above; hence $\sum_{j=1}^k a^2_j = \sum_{j=1}^k b^4_j$ has a maximum value. Let $l$-numbers $(1 \le l \le k)$ of $b_j$ be non-zero; $b_{l+1} = \cdots = b_k =0$; then it follows that
\begin{equation}\label{eqs:keiyu}
2 b^2_j +  \gamma  = 0 \ ( j = 1,\ldots,l). 
\end{equation}
Summing these equations over $j=1,\ldots,l$ and considering Eq.~\eqref{eq:5}, one obtains 
\begin{equation}\label{eq:gamma}
\gamma = - \frac{2}{l} (\sum_{j=k+1}^N |a_j|), 
\end{equation}
and  
\begin{equation}\label{eq:extr}
\sum_{j=1}^k a^2_j = \sum_{j=1}^k b^4_j = \frac{1}{l} (\sum_{j=k+1}^N |a_j|)^2,
\end{equation}
where $b_j$s $ (j = 1,\ldots,k)$ are 
\begin{equation}\label{eq:solofa_j}
b^2_j = \frac{1}{l} (\sum_{j=k+1}^N |a_j|) \ (j=1,\ldots,l), \ b_{l+1} = \cdots  = b_{k} = 0, 
\end{equation}
from Eqs~\eqref{eqs:keiyu} and Eq.~\eqref{eq:gamma}. Hence it takes the maximum value at $l=1$: 
\begin{equation}\label{eq:max}
\sum_{j=1}^k a^2_j = (\sum_{j=k+1}^N |a_j|)^2,
\end{equation}
where only $b_1$ is non-zero:  
\begin{equation}\label{eq:solofa_j}
b^2_1 = \sum_{j=k+1}^N |a_j|,  
\end{equation}
with $b_{2} = \cdots  = b_{k} = 0$. From assumption \eqref{eq:assum} and Eq.~\eqref{eq:max}, it is bounded from above as
\begin{equation}\label{eq:ineqcontara2}
\sum_{j=1}^k a^2_j < \frac{2(N-k)^2}{N(N-1)}. 
\end{equation}
On the other hand, the inequality  
\begin{equation}\label{eq:refineq}
2 - 2 \frac{N-k}{N(N-1)} \ge \frac{2(N-k)^2}{N(N-1)} \Leftrightarrow (k-1) (2N-k) \ge 0 
\end{equation}
holds since $1\le  k \le N-1$, and hence $2N -k > 0$. Consequently, Eq.~\eqref{eq:ineqcontara2} is contradictory to Eq.~\eqref{eq:ineqcontra}; namely, $|m(\lambda_{\bm n})| \ge \sqrt{\frac{2}{N(N-1)}}$. This completes the proof of [I].  

Notice that, assuming the equality $|m(\lambda_{\bm n})| = \sqrt{\frac{2}{N(N-1)}}$ holds, the above discussion also contains the proof of [III] if one follows the logic changing $< (>)$ into $\le (\ge)$, respectively, in Eqs.~\eqref{eq:assum}, \eqref{eq:ineqcontra}, and \eqref{eq:ineqcontara2}. Only when $k=1$ are there solutions (see Eq.~\eqref{eq:refineq}) provided that all $a_2 = \cdots = a_N = - \sqrt{\frac{2}{N(N-1)}}$ (notice that equalities in Eqs.~\eqref{eq:assum} must hold for the equality in Eq.~\eqref{eq:ineqcontra} after changing $< (>)$ into $\le (\ge)$); and $M(\lambda_{\bm n} = a_1) = \sqrt{\frac{2}{N(N-1)}}$ from Eq.~\eqref{eq:solofa_j}. This completes the proof of [III].  

{\hfill QED}

Applying these properties to Theorem \ref{Thm:Main}, first we notice that [I] in Proposition \ref{Prop:EiSUN} reproduces the range \eqref{eq:restrictionofbmax} (or inclusion relation \eqref{eq:includion}); furthermore, we obtain a dual property of the Bloch-vector space: 
\begin{Thm}\label{eq:Dual}
The Bloch-vector space has the following dual property. 

[A] In the Bloch-vector space, if in some direction where the space goes to the surface of large ball $D_{r_{\max}}(\R^{N^2-1})$, i.e., if there is a pure state in that direction, in the opposite side the space can only reach the surface of small ball $D_{r_{min}}(\R^{N^2-1})$. 

[B] Conversely, if in some direction the space only reaches the surface of the small ball, then in the opposite side the space can reach the surface of the large ball. 

\end{Thm}
In other words, if in some direction there is a fold sticking out into a pure state, then the opposite side should be concave \cite{fn:4}, and vice versa. We expect this geometrical view of Theorem \ref{eq:Dual} will give us an overall picture of the Bloch-vector space.

{\bf Proof of Theorem \ref{eq:Dual}}

 [A] and [B] immediately follow from [III] and [II], respectively, in Proposition \ref{Prop:EiSUN} since $m(\lambda_{-{\bm n}}) = - M(\lambda_{\bm n})$. ([A] might be obtained more easily in another way; we give one of the independent proofs for the reader's convenience in Appendix \ref{app:IndProof}). 

{\hfill QED}

From [II] and [III], $|m| = \sqrt{\frac{2}{N(N-1)}}$ or $\sqrt{\frac{2(N-1)}{N}}$ implies $M = \sqrt{\frac{2(N-1)}{N}}$ or $\sqrt{\frac{2}{N(N-1)}}$, respectively; hence it holds that $\frac{2}{N|m|}= M$. Consequently we also obtain
\begin{Prop}\label{Prop:MaxDet1}
The Bloch-vector space is determined by the maximal eigenvalue (see Eq.~\eqref{eq:BisSubsetByM}) if the space gets to the surface of the large ball, or only reaches the surface of the small ball in the corresponding direction. 
\end{Prop}  
In $2$-level and $3$-level systems, the opposite is also true:  
\begin{Prop}\label{Prop:MaxDet2}
The Bloch-vector spaces in $2$-level and $3$-levele systems are determined by maximal eigenvalue if and only if the space gets to the surface of the large ball or only reaches the surface of the small ball in the corresponding direction.  
\end{Prop}  

{\bf Proof }
The case for a $2$-level system is trivial since the small and large ball coincides. Let us consider a $3$-level system. Assume that eigenvalues of $\lambda_{{\bm n}}$ are $a_1 = \frac{2}{3 |m|} \ge  a_2 \ge a_3 = - |m|$. Then it is straightforward to see the solutions of Eq.~\eqref{eqs:lambdan2} are $|m| =\sqrt{\frac{2}{3(3-1)}}$ or $\sqrt{\frac{2(3-1)}{3}}$, which are the case where in the direction the space gets to the surface of the large ball or only reaches the surface of the small ball.   
{\hfill QED} 

Propositions \ref{Prop:MaxDet1} and \ref{Prop:MaxDet2} are also seen in Fig.~\ref{fig:i8}. At the angle $\theta = \pi/6, \ 5\pi/6, \ 3\pi/2$ (pure states) and $\theta = 7\pi/6, 11\pi/6, \pi/2 $ (small ball), one sees that the Bloch-vector space is determined by the maximal eigenvalues which are plotted by a broken line.

On the other hand, for higher-level systems, i.e., $N \ge 4$, the opposite does not necessary hold. For example, in $4$-level systems, there exists some direction ${\bm n}$ where corresponding generators have a matrix representation of 
\begin{equation}\label{eq:counterEx}
\lambda({\bm n}) =
 \frac{1}{\sqrt{2}}
 \left(
 \begin{array}{cccc}
 0&1&0&0\\
 1&0&0&0\\
 0&0&0&-i\\
 0&0&i&0\\
 \end{array}
 \right).
\end{equation}
Since the eigenvalues are $\frac{1}{\sqrt{2}},-\frac{1}{\sqrt{2}}$, it holds that $\frac{2}{4 |m|} = M \ (m = -\frac{1}{\sqrt{2}}, M = \frac{1}{\sqrt{2}})$, while it does not equal $\sqrt{\frac{2}{4(4-1)}}$ nor $\sqrt{\frac{2(4-1)}{4}}$. This can be seen in B) in Fig.~\ref{fig:34level} in Appendix \ref{2dimSec}.    

\section{Parameterization of density operators}\label{sec:para}

In this section, we show another application of Theorem \ref{Thm:Main}, namely parameterization of quantum states. It is often helpful to generate quantum states numerically if one wants to investigate some global properties of quantum state by numerical experiments. In order to do that, one needs parameterization of quantum states with real values, especially whose range of parameter should be simple enough to make the numerical computations easier. 

From Theorem \ref{Thm:Main}, the Bloch vector in spherical-coordinate provides a natural parameterization of the quantum state (density matrix): the direction vector ${\bm n}$ can be parameterized with $(N^2-2)$-numbers of angles, e.g.,  
\begin{equation}
n_i = (\prod_{j=1}^{i-1}\sin \theta_j) \cos \theta_i, \ (i=1,\ldots,N^2-3),\quad n_{N^2-1} = (\prod_{j=1}^{N^2-3}\sin \theta_j) \sin \theta_{N^2-2},
\end{equation}  
with ranges 
\begin{equation}\label{eq:range1}
 0 \le \theta_i \le \pi, (i=1,\ldots,N^2-3),\quad 0 \le \theta_{N^2-2} < 2 \pi, \end{equation}
while the range for radius $r$ is  
\begin{equation}\label{eq:range2}
0 \le r \le \frac{2}{N|m(\lambda_{\bm n})|}. 
\end{equation}
This parameterization is helpful when one wants to generate all the density operators systematically, since the ranges \eqref{eq:range1} and \eqref{eq:range2} of the parameters are simple enough.

\section{Class of quantum-state representations based on expectation values}

So far we have restricted our attention to the Bloch-vector as one of the quantum-state representations based on actual measurements; the minimum required set of observables to determine quantum state are those of orthogonal generators of $SU(N)$. However, such a set is not uniquely determined, and there seem to be no reasons to restrict to the set. In fact, one should be cautious enough in choosing a set of observables for the representation of states, since each choice inevitably violates equality among observables by specifying the set of observables. Therefore, it is interesting to classify such representations according to the physical and mathematical backgrounds of the choice of the set of observables. 

In this section, we provide three classes of quantum-state representation based on expectation values (C1), (C2), and (C3), with hierarchy (C1) $\supset$ (C2) $\supset$ (C3), where class (C3) comes to that of the Bloch vector; hence (C1) and (C2) include more general representations. We also note that the class (C1) includes quantum state even with infinite levels, while the representation by the Bloch-vector (C3) inevitably limits the case to the quantum systems with finite levels, since the unit operator is not a trace class one. 

To start from a general setting, let $\HA$ be an associated (separable) Hilbert space to a quantum system where the dimension of $\HA$ is not necessarily finite. Then, quantum state is usually represented by a density operator $\rho$, which is a trace class operator, hence also a Hilbert-Schmidt one, with conditions $\rho \ge 0$ and $\tr \rho = 1$. 

\noindent (C1) {\bf Quantum-state representation with expectation values of dual basis}  

One of the essences in representing quantum state by the Bloch vector is to span the density operator by linear combinations of observables. However, for the set of observables, one can use an arbitrary basis of the self-adjoint Hilbert-Schmidt class $\C_2(\HA)$ since density operators are elements there. Note that the set $\C_2(\HA)$ is a real (separable) Hilbert space with the inner product $(A,B) = \tr A B$. With an observable $A = A^\dagger$ and a density operator $\rho$, the inner product is nothing but its expectation value: $(\rho,A) = \tr \rho A = \langle A \rangle$. Let $\{A_i = A^\dagger_i\}_{i=1}^{D}$ ($D \equiv $ dim $(\C_2(\HA)) \le \infty $) be a basis not necessarily orthonormal, with its dual basis $\{\tilde{A}_i = \tilde{A}^\dagger_i\}_{i=1}^{D}$ \cite{fn:5}:  
\begin{equation} 
(\tilde{A}_i,A_j) = \tr \tilde{A}_iA_j = \delta_{ij}. 
\end{equation}
Then, any density operator $\rho$ can be written in the form 
\begin{equation}\label{eq:expansion}
\rho = \sum_{i=1}^D (\tilde{A}_i, \rho) A_i = \sum_{i=1}^D \langle \tilde{A}_i\rangle A_i.\end{equation}
This expansion implies that expectation values of dual basis $\langle \tilde{A}_i\rangle$ determine the density operator through this equation. Therefore, one can consider that ${\bm a} \equiv (\langle \tilde{A}_1\rangle,\langle \tilde{A}_2\rangle,\ldots)$ itself represents a quantum state equivalently to the density operator. Since there is an arbitrariness in choosing observables $\{\tilde{A}_i \}$ as a dual basis, each choice provides different representation and we define the class (C1) as that which is composed of such representations.

\noindent (C2) {\bf Quantum-state representation with expectation values of dual basis including normalization condition}  

Different from the representation by the Bloch vector, normalization condition $\tr \rho = 1$ has not yet considered in class (C1). This can be done by choosing a unit operator as one of the elements of the basis. However, since the unit operator is not in $\C_2(\HA_N)$ in infinite-dimensional cases, one needs to restrict oneself to quantum systems with finite levels (dim $\HA = N < \infty$) again. 

The next class (C2), included in (C1), is the representation for $N$-level systems where $\I_N$ is used as one of the elements of the basis $A_i$s --- let $A_{N^2} = \I_N$ --- and others $A_i \ (i=1,\ldots,N^2-1)$ being orthogonal to $\I_N$, i.e., $(\I_N, A_i) = \tr A_i = 0 \ (i=1,\ldots,N^2-1)$. Then it holds that $\tilde{A}_{N^2} = \frac{1}{N}\I_N$, while $\tilde{A}_i \ (i=1,\ldots,N^2-1)$ are also self-adjoint traceless operators; hence $A_i$s and $\tilde{A}_i$s are generators of $SU(N)$ (cf. Appendix \ref{sec:GeneratorsOsSUN}). Using these bases, Eq.~\eqref{eq:expansion} becomes 
\begin{equation}
\rho = (\tr \frac{1}{N}\I_N \rho) \I_N + \sum_{i=1}^{N^2-1}\langle \tilde{A}_i\rangle A_i = \frac{1}{N} \I_N + \sum_{i=1}^{N^2-1}\langle \tilde{A}_i\rangle A_i,\end{equation}  
where the normalization condition $\tr \rho = 1$ automatically satisfies in this form. From this equation, one can consider $(N^2-1)$-dimensional real vector ${\bm a} \equiv (\langle \tilde{A}_1\rangle,\ldots,\langle \tilde{A}_{N^2-1}\rangle)$ itself a representation of quantum state where normalization condition is naturally included from the beginning. The class (C2) of the quantum-state representation is that composed of such representations.

\noindent (C3) {\bf Quantum-state representation by the Bloch vector}  

Now it is obvious that the representation by the Bloch vector is the special case in class (C2) where the set of observables are orthogonal generators of $SU(N)$ \cite{fn:factor}. Namely, we require also orthogonality condition between $A_i$s. Since there are still as many orthogonal generators of $SU(N)$ as there are elements of $O(N^2-1)$, each choice provides different Bloch-vector representation, and we call this class (C3).

 Although the orthogonality condition among $A_i$s might make some calculations easy, it seems that this is a purely mathematical condition without particular physical meaning. Even when one considers only this point, it is interesting to proceed to (C1) and (C2) representations beyond (C3) of the Bloch-vector representation. 
 
\vspace{2mm}

\noindent {\bf State space for classes (C1) and (C2)}  

\vspace{2mm}

When introducing state-representation, it is of importance to discuss its state-space. In particular, when understanding its geometrical character, it is helpful to comprehend global information of not only states but also dynamics.  

We note that almost the same discussion as in Sec.~\ref{sec:RadiusOfBS} using the spherical-coordinate view can be applied to the classes (C1) and (C2). It is interesting to note that for the state space of the class (C2), the same characterization as that of Bloch vectors (Theorem \ref{Thm:Main}) holds when changing $\lambda_{\bm n}$ to $A_{\bm n}$ and removes factor $2$, since the orthogonality condition (iii) is not used in that proof. As we know that the Bloch-vector space has a quite complex structure except in a $2$-level system, it would be interesting to investigate another representation in class (C2) by finding some generators of $SU(N)$ whose minimum eigenvalues can be characterized with simple conditions; then from Theorem \ref{Thm:Main} the state-space would be simpler than that of the Bloch vector.

\section{Conclusion and discussion}\label{sec:dis}

We have discussed the Bloch-vector space from the spherical-coordinate point of view and showed that the radius is determined not by the maximum but the minimum eigenvalue of generators $\lambda_{\bm n}$ in each direction (Theorem \ref{Thm:Main}). Compared with the analytic determination in Ref.~\cite{ref:Gen}, Theorem \ref{Thm:Main} brings us to understand some geometrical characters in the complex structure of the space, like Theorem \ref{eq:Dual}, Proposition \ref{Prop:MaxDet1}, and Proposition \ref{Prop:MaxDet2}: When there is a prominence which goes to the large ball, namely to some pure state, then in opposite side the space can only reach the small ball, and vice versa, etc. These give us an overall picture of the Bloch-vector space, even in higher-level cases. The physical meaning of this theorem is also interesting; since the radius $r$ of the Bloch vector is related to the purity $P(\rho) \equiv {\mathrm tr} \rho^2$ by $P(\rho) = \frac{(1 + r^2)}{N}$, if there is a pure state in some direction, in the opposite direction one has to give up the high purity. On the other hand, if in some direction the purity is low enough, then in the opposite direction one can get high purity. The knowledge of the geometry of the state-space provides also other applications, like a positive map \cite{ref:Kossakowski,ref:Gen&Koss}, which can be used as a dynamical map, or even a tool to distinguish separable and entanglement \cite{ref:Hor}.   

We also introduced classes of quantum-state representations based on actual measurements, beyond the description by the Bloch vector. In particular, the state space of class (C2) can be characterized in the same manner as Theorem \ref{Thm:Main}. We believe that there exists more useful representation than that of the Bloch vector in the sense that the state-space has simpler structure, and Theorem \ref{Thm:Main} would be useful for the search of such representation since it tells us that all the information of the state-space lies in minimum eigenvalues of observable in each direction.

\acknowledgments

G.K. gratefully acknowledges Prof.~I. Ohba, Prof.~H. Nakazato, Prof.~S. Tasaki, Prof.~S. Pascazio, Prof.~A. Miranowicz, Dr. K. Imafuku, and Dr. M. Miyamoto for their helpful discussions and fruitful comments. In particular, he is grateful to Prof.~K. \.Zyczkowski for reading the manuscript prior to publication and making fruitful comments. This research is partially supported by the Grant-in-Aid for JSPS Research Fellows and the Grant PBZ-MIN-008/P03/2003. 

\appendix

\section{The generators of $SU(N)$}\label{sec:GeneratorsOsSUN}

The definition and some important properties of the generators of $SU(N)$ are reviewed here. Generators of $SU(N)$ are defined by $(N^2-1)$-numbers of independent operators $\lambda_i \ (i=1,\ldots,N^2-1)$ on $\HA_N$ which satisfies 
\begin{equation}\label{eq:generatorsg}
\mbox{(i)} \ \lambda^\dagger_i = \lambda_i, \mbox{(ii)} \ \tr \lambda_i = 0, 
\end{equation}
with which any $U \in SU(N)$ is given by $U = \exp (i \sum_{i=1}^{N^2-1} \alpha_i \lambda_i) \ (\alpha_i \in \R)$. Usually, the orthogonal condition (in the sense of Hilbert-Schmidt inner product) i.e.,  (iii) $\tr \lambda_i\lambda_j = 2\delta_{ij}$ is also required for simplicity. Note that the factor $2$ is due to convention and there are no logical reasons to use it \cite{fn:norfac}. We will call them orthogonal generators of $SU(N)$ which satisfy  
\begin{equation}\label{eq:generators}
\mbox{(i)} \ \lambda^\dagger_i = \lambda_i, \mbox{(ii)} \ \tr \lambda_i = 0, \mbox{(iii)} \ \tr \lambda_i\lambda_j = 2\delta_{ij}.  
\end{equation}
Physically, the condition (i) implies that one can consider them some observables. With identity operator $\I_N$, they form a complete orthogonal basis for the set of all linear operators on $\HA_N$. (Trivially the dimension is $N^2$; the condition (ii) means the orthogonality between the identity operator $\I_N$ between all $\lambda_i$s; and the condition (iii) also means those among $\lambda_i$s). Hence any operator including density operator can be expanded by them as in Eq.~\eqref{eq:Exprho} and this leads to the notion of the Bloch vector. 

Since $-i[\lambda_i,\lambda_j]$ and $[\lambda_i,\lambda_j]_+$, where $[\cdot,\cdot]$ and $[\cdot,\cdot]_+$ are commuting and anti-commuting relations respectively, are Hermitian operators, they can be expanded by $\{\I_N, \lambda_i \}$ with real coefficients; furthermore considering the properties \eqref{eq:generators}, they can be written in the form
\begin{equation}\label{eq:comanticomofgen}
-i[\lambda_i,\lambda_j] = 2 \sum_{k=1}^{N^2-1} f_{ijk} \lambda_k, \quad  [\lambda_i,\lambda_j ]_+ = \frac{4}{N} \I_N +  2 \sum_{k=1}^{N^2-1} g_{ijk} \lambda_k, \end{equation}
with some constants $f_{ijk}, g_{ijk} \in \R$, called structure constants. Adding these equations together, the multiplication of orthogonal generators are given by
\begin{equation}\label{eq:multilambda}
\lambda_i\lambda_j = \frac{2}{N}\I_N + \sum_{k=1}^{N^2-1}z_{ijk}\lambda_k,
\end{equation}
where $z_{ijk} \equiv g_{ijk} + i f_{ijk}$ is the complex structure constant. By multiplying $\lambda_k$ and taking traces in Eqs.~\eqref{eq:comanticomofgen}, structure constants can be rewritten as 
\begin{equation}\label{eq:StructConst}
f_{ijk} \equiv \frac{1}{4i} \tr [\lambda_i,\lambda_j]\lambda_k,\ g_{ijk} = \frac{1}{4} \tr [\lambda_i,\lambda_j]_+ \lambda_k. 
\end{equation}
From this, it is easy to see $f_{ijk}$ and $g_{ijk}$ are completely antisymmetric and symmetric in the displacement of any pair of indices.

It is important to note that orthogonal generators \eqref{eq:generators} are not uniquely determined; in stead of that, any possible generators which satisfies Eqs.~\eqref{eq:generators} are connected by $[O_{ij}] \in O(N^2-1) $ of orthogonal group: Let $\{\lambda_i \}_{i=1}^{N^2-1}$ and $\{\lambda^\prime_j \}_{j=1}^{N^2-1}$ orthogonal generators of $SU(N)$, then there exists $[O_{ij}] \in O(N^2-1) $ such that
\begin{equation}\label{eq:NonUniqGen}
\lambda^\prime_i = \sum_{j=1}^{N^2-1}O_{ij} \lambda_j. 
\end{equation}
Conversely, with some generators $\{\lambda_i \}_{i=1}^{N^2-1}$ and any orthogonal matrix $[O_{ij}] \in O(N^2-1) $, $\{\lambda^\prime_i \}_{i=1}^{N^2-1}$ in Eq.~\eqref{eq:NonUniqGen} are orthogonal generators of $SU(N)$. Note that, $\lambda_{\bm n}$ for any direction ${\bm n}$ can be considered as an element of some generators of $SU(N)$; e.g., by choosing orthogonal matrix $[O_{ij}]$ as $n_i = O_{ij}n^\prime_j $ where ${\bm n^\prime} = (0,0,\ldots,0,1)$, then $\lambda_{\bm n}$ is $(N^2-1)$ th element of generators $\lambda^\prime_i \equiv O_{ij}\lambda_j$: $\lambda^\prime_{N^2-1} = \lambda_{\bm n}$. We stress that for this reason general properties for generators of $SU(N)$ are also applied to $\lambda_{\bm n}$ in any direction, and vice versa. 


\section{Another proof for [A] in the Theorem \ref{eq:Dual}}\label{app:IndProof}

We give another proof for one of the dual properties [A] in Theorem \ref{eq:Dual}: Let $\rho$ be pure state, then it follows $\rho^2 = \rho$. Inserting $\rho = \frac{1}{N}\I_N + \frac{1}{2}r\lambda_{{\bm n}}$, one obtains
\begin{equation}
\frac{r^2}{4} \lambda_{{\bm n}}^2 +  \frac{r(2-N)}{2N} \lambda_{{\bm n}} + \frac{1-N}{N^2}\I_N = 0. 
\end{equation}
Since the radius for pure state is $r = r_l= \sqrt{\frac{2(N-1)}{N}}$, it follows  
\begin{eqnarray}
\frac{N -1}{2N} \lambda_{{\bm n}}^2 +  \sqrt{\frac{2(N-1)}{N}}\frac{2-N}{2N}  \lambda_{{\bm n}} + \frac{1-N}{N^2}\I_N = 0 \nonumber \\
\Leftrightarrow \frac{N -1}{2N}(\lambda_{{\bm n}} - \sqrt{\frac{2(N-1)}{N}})(\lambda_{{\bm n}} + \sqrt{\frac{2}{N(N-1)}}) = 0. 
\end{eqnarray}
Namely, the eigenvalues of $\lambda_{{\bm n}}$ are $\sqrt{\frac{2(N-1)}{N}}$ and $ - \sqrt{\frac{2}{N(N-1)}}$. Consequently the minimum eigenvalue of opposite direction $m(\lambda_{-{\bm n}}) = -\sqrt{\frac{2(N-1)}{N}}$ and maximum radius $\frac{2}{N|m(\lambda_{-{\bm n}})|}$ is that of small ball.  

\hfill{QED}

\section{$2$-dimensional section of the Bloch-vector space}\label{2dimSec}

To visualize Theorem \ref{Thm:Main}, Theorem \ref{eq:Dual}, Proposition \ref{Prop:MaxDet1}, and Proposition \ref{Prop:MaxDet2}, we provide $2$-dimensional sections of the Bloch-vector space for $3$-level \cite{ref:Mahler,ref:Gen} and $4$-level \cite{ref:Jakobczyk} systems, where $2$-dimensional sections $\Sigma_2(i,j)$ \cite{ref:Jakobczyk,ref:Mahler} are defined as $\Sigma_2(i,j) = \{ {\bm b} \in B({\mathbb R}^{N^2-1}) : {\bm b} = (0,\ldots,0,b_i,0,\ldots,0,b_j,0,\ldots,0)\}$. 

\begin{figure}    
\includegraphics[height=1.0\textwidth]{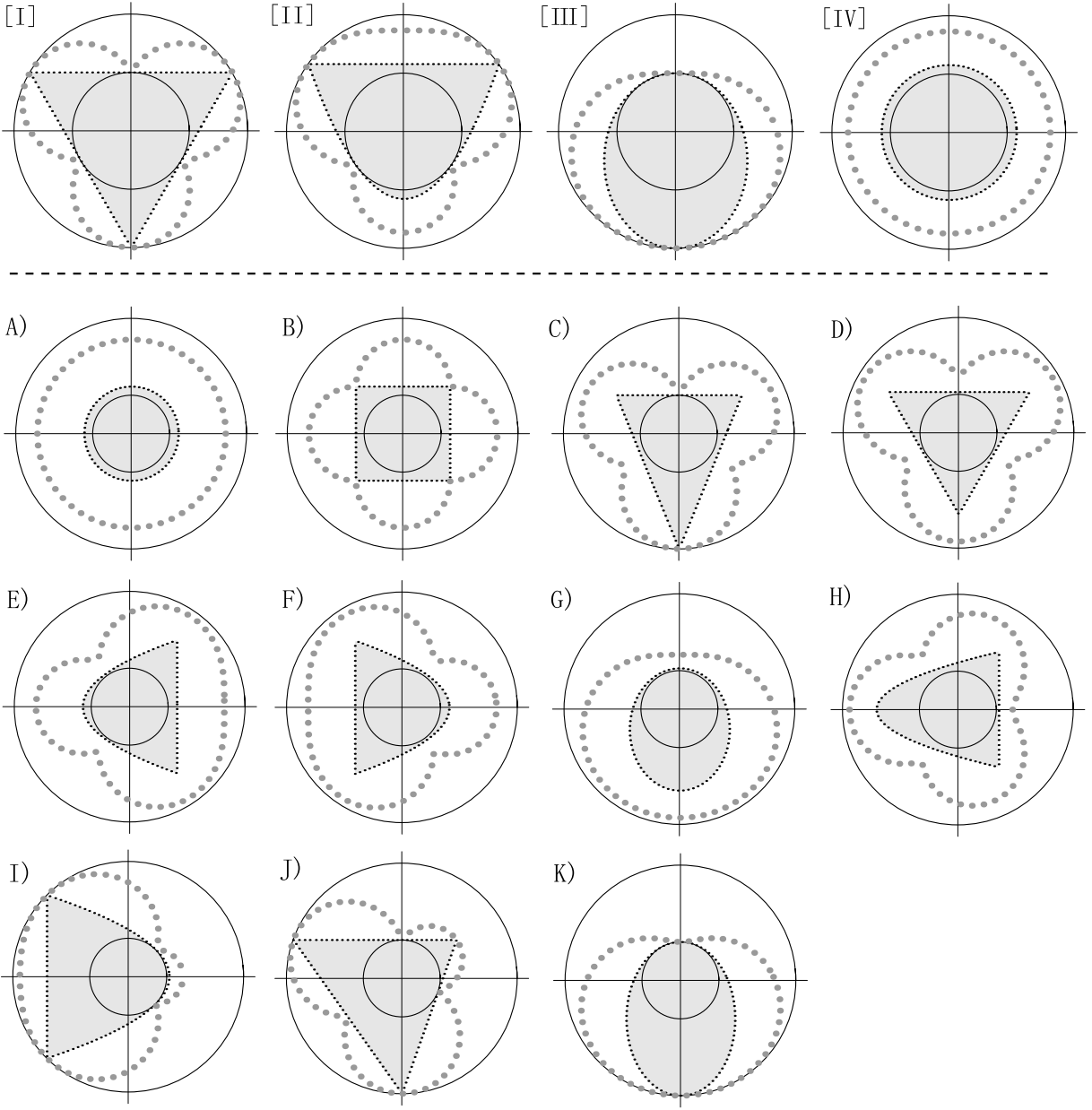}
\caption{$2$-dimensional sections of the Bloch-vector space for $3$-level systems (from [I] to [IV]) and $4$-level systems ( from A) to K) ). As for the classification of the sections, see the Refs.~\cite{ref:Gen} and \cite{ref:Jakobczyk} for $3$-level systems and $4$-level systems respectively.  }\label{fig:34level}
\end{figure}

In Fig.~\ref{fig:34level}, all possible $2$-dimensioanl sections for $3$-level and $4$-level systems are plotted in accordance with a classification in Ref.~\cite{ref:Gen} for $3$-level systems from [I] to [IV]; in Ref.~\cite{ref:Jakobczyk} for 4-level systems from A) to K) fixing special orthogonal generators (See each reference for the details), which are plotted by numerically solving eigenvalue-problems for generators and using Theorem \ref{Thm:Main}.  

Like in figure \ref{fig:i8}, all the grey regions are $2$-dimensional sections of the Bloch-vector space with sections of large and small balls $ D_{r_{l}}(\R^{N^2-1}) $ and $D_{r_{s}}(\R^{N^2-1})$ for $N=3$ from [I] to [IV] and $N=4$ from A) to K) ( Cf. inclusion relations \eqref{eq:includion}). The maximum eigenvalues are also plotted in dotted line, which shows that they do not necessary determine the Bloch-vector space. The dual properties (Theorem \ref{eq:Dual}) are seen in figures [I], [II], [III] in $3$-level case, and also in figures C), I), J), and K) in $4$-level case, where one also notices Proposition \ref{Prop:MaxDet1}, and Proposition \ref{Prop:MaxDet2}. In figure B), one sees that maximum eigenvalue determines the Bloch-vector space not on the small nor large ball; which was in fact used in Eq.~\eqref{eq:counterEx}.   

\begin{thebibliography}{00}
\bibitem{fn:1}
Since pure state $\rho$ satisfies $\tr \rho^2 = 1$, one immediately sees that the corresponding Bloch vector is on the ball by inserting Eq.~\eqref{eq:mapB&Rho} into $\tr \rho^2$ and using Eqs.~\eqref{eq:generators0}
\bibitem{fn:zyc}
The volume of the mixed states is comupted comparing to the large and small ball in Ref.~\cite{ref:Zyczkowski}.
\bibitem{fn:2}
A radius $r \ge 0$ is also bounded below by the minimum eigenvalue $m(\lambda_{\bm n})$. However, as is shown in Sec.~\ref{sec:RadiusOfBS}, $m(\lambda_{\bm n})$ is less than $0$ and hence the lower bound is trivial.
\bibitem{fn:3}
Note that $\frac{1}{N}\I_N$ is a density operator; maximally mixed state.
\bibitem{fn:4}
This expression might give a false impression: Note that the space has the property of convexity.
\bibitem{fn:5}
Notice that if $D < \infty$ there always exists dual basis for any basis $\{A\}_{i=1}^D$: If one defines linear functional $f_i: \C_2 \to \R$ by $f_i(A_j) \equiv \delta_{ij}$, then it is obviously bounded subject to $D < \infty$. Since $\C_2$ forms a Hilber space, there uniquely exists $\tilde{A}_i \in \C_2$ such that $f_i(B) = \tr \tilde{A}_i B$ for any $B\in \C_2$ by Riesz representation theorem \cite{ref:Reed&Simon}.
\bibitem{fn:factor}
To be precise, the representation here should be divided by the factor $2$, since we used conventional normalization for orthogonal generators of $SU(N)$.
\bibitem{fn:norfac} 
For instance, if one takes this factor not $2$ but $N$ in $N$-dimensional cases, the results in Theorem \ref{Thm:Main} changes from $||{\bm b}_{\bm n}||_{\max} = \frac{2}{N|m(\lambda_{\bm n})|}$ to $||{\bm b}_{\bm n}||_{\max} = \frac{1}{|m(\lambda_{\bm n})|}$, which looks simpler in cosmetic point of view.
\end{thebibliography}
\end{document}